\documentclass[a4paper, twocolumn]{revtex4-2} 
\usepackage{color}
\usepackage{graphics}
\usepackage{epsfig}

\begin{document}
	\title{Planar Hall effect in Cu intercalated PdTe$_2$}
	
	\author{Sonika$^1$, M. K. Hooda$^{1,2}$, Shailja Sharma$^1$ and C. S. Yadav$^{1 *}$}
	
	\affiliation{$^1$ School of Basic Sciences, Indian Institute of Technology Mandi, Kamand Mandi-175075 (H.P.) India}
	\affiliation{$^2$Department of Physics, Indian Institute of Technology Kanpur, Kanpur-208016 (U.P.), India}
	
	
	\begin{abstract}
		
		We present the Planar Hall effect studies on the Cu intercalated type-II Dirac semimetal PdTe$_{2}$. The electrical resistivity exhibits a positive field dependence both in perpendicular and parallel field directions, causing non-zero anisotropy. The longitudinal magnetoresistance shows almost linear field dependence at low temperatures. A tilted prolate spheroid shaped orbits are observed in parametric plot between transverse and longitudinal resistivities. Our study suggest that for the type-II Dirac semimetal materials with positive longitudinal magnetoresistance, the origin of Planar Hall effect cannot be asserted with certainty to the topological or non-topological without taking into account the anisotropy of Fermi surface. 	
	\end{abstract}
	
	\maketitle
	
	\section{Introduction}
	
	Topological (Dirac/Weyl) semimetals have attracted considerable attention because of their exotic physical properties like high carrier mobility, large positive magnetoresistance (MR), anomalous Hall effect, anomalous Nernst signals \textit{etc.} \cite{liang2015ultrahigh, shekhar2015extremely, ali2014large, kumar2017extremely, liang2018anomalous, ikhlas2017large}. One of the salient features of topological semimetals is the chiral or Adler-Bell-Jackiw anomaly \cite{liang2017anomalous, adler1969axial, bell1969pcac, kim2013dirac}. Chiral anomaly pumps the charge carriers between two Weyl points that induces charge current when magnetic field is applied parallel to the electric field. This effect results in negative longitudinal MR. Therefore, negative longitudinal MR has been used as a signature to identify chiral anomaly in a system \cite{xiong2015evidence, huang2015observation, li2016negative, hirschberger2016chiral}. However, several other factors like current jetting, weak localization, conductivity fluctuations, suppression of magnetic scattering and impurity induced scattering, may also contribute to negative longitudinal MR \cite{hu2005current, arnold2016negative, ulmet1988negative, dos2016search, zhang2017tunable, schumann2017negative, goswami2015axial}.
	
	After the reports of Planar Hall effect (PHE) in Weyl metals by Burukov \textit{et al.} \cite{burkov2017giant}, Nandi \textit{et al.} showed that chiral anomaly is responsible for PHE in Weyl semimetals \cite{nandy2017chiral}. Following these theoretical predictions, PHE was studied in topological semimetals and used as a probe to identify chiral anomaly. Unlike usual Hall effect, PHE manifests itself when magnetic field (\textit{H}) is rotating in the same plane as that of current (\textit{I}) and induced Hall voltage. PHE has been observed in the materials like ferromagnetic metals, topological insulators and topological semimetals. The PHE in ferromagnetic metals generally have small magnitudes and is caused due to interaction between magnetic order and spin-orbit coupling \cite{tang2003giant, nazmul2008planar}. Breaking of time reversal symmetry in topological insulators gives birth to PHE in compounds like Bi$_2$Se$_3$ and Bi$_{2-x}$Sb$_x$Te$_3$ \cite{he2019nonlinear, taskin2017planar}. Theoretical predictions fits well for many topological semimetals like ZrTe$_5$, Na$_3$Bi, GdPtBi, WTe$_2$, Cd$_3$As$_2$, VAl$_3$, MoTe$_2$, Co$_3$Sn$_2$S$_2$, PdTe$_2$ \textit{etc.} \cite{li2018giant, liang2018experimental, kumar2018planar, li2019anisotropic, wang2018planar, wu2018probing, singha2018planar, chen2018planar, SHAMA2020166547, xu2018planar}. However, Kumar \textit{et al.} highlighted the fact that orbital MR always exist in real systems and cannot be neglected \cite{kumar2018planar}. Orbital MR is the enhancement of resistivity due to Lorentz force experienced by charged carriers on the application of magnetic field. Later the anisotropy in orbital MR was considered as the origin of PHE in compounds like  NiTe$_2$, PtSe$_2$, TaP, PdTe$_2$ \textit{etc.} \cite{liu2019nontopological, li2020planar, yang2019current, meng2019planar}. The co-existence of orbital MR and the chiral anomaly raises a question `whether PHE can be used as a diagnostic tool to detect chiral anomaly in a system or not?' 
	
	PdTe$_2$ is a type-II Dirac semimetal which shows positive MR \cite{noh2017experimental, fei2017nontrivial}. Xu \textit{et al.} first reported the PHE in PdTe$_2$ and attributed its origin to the chiral anomaly \cite{xu2018planar}. However, Meng \textit{et al.} claimed that anisotropic orbital MR is responsible for PHE in PdTe$_2$ \cite{meng2019planar}. These observations stimulated our interest in studying PHE in the corresponding intercalated compound Cu$_{0.05}$PdTe$_2$ to understand the origin and effect of intercalation on the amplitude of PHE. The method of preparation of single crystalline samples of Cu$_{0.05}$PdTe$_2$ has been reported elsewhere \cite{PhysRevB.103.235105}. All the measurements were performed using PPMS (Dynacool) by Quantum Design Inc. 
	
	We have measured angle dependent PHE and anisotropic longitudinal resistivity in Cu$_{0.05}$PdTe$_2$. Our observation of positive longitudinal magnetoresistance (LMR), linear field dependence of the amplitude of PHE and the prolate shaped orbits in parametric plot point towards the importance of Fermi surface anisotropies (FSA) in understanding the origin of PHE in a system like PdTe$_2$.
	
	\section{Results and Discussion}
	
\begin{figure}[tb]
		\includegraphics[width=8 cm]{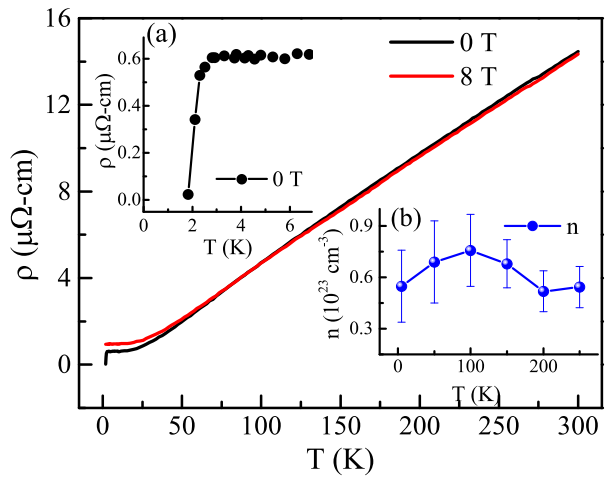}
		\caption{Electrical resistivity of Cu$_{0.05}$PdTe$_2$ at \textit{H} = 0 and 8 T. The inset (a) shows the superconducting transition at $\sim$2.5 K (b) shows the carrier concentration of Cu$_{0.05}$PdTe$_2$.}
		\label{fig:Figure1}
	\end{figure}

	
	Figure \ref{fig:Figure1} shows the temperature dependence of electrical resistivity ($\rho$) of Cu$_{0.05}$PdTe$_2$. The compound shows superconducting transition at $\sim$2.5 K in agreement with literature (Inset (a) of fig. \ref{fig:Figure1}) \cite{ryu2015superconductivity, hooda2018electronic, vasdev2019enhanced}. The transition disappears on the application of magnetic field and $\rho(\textit{T})$ shows metallic nature. A high carrier concentration (\textit{n}) of the order 10$^{22}$ cm$^{-3}$ with \textit{p}-type charge carriers has been observed in Hall measurement (Inset (b) of fig. \ref{fig:Figure1}).
	
	\begin{figure}[tb]
		\includegraphics[width=7 cm, height=10cm]{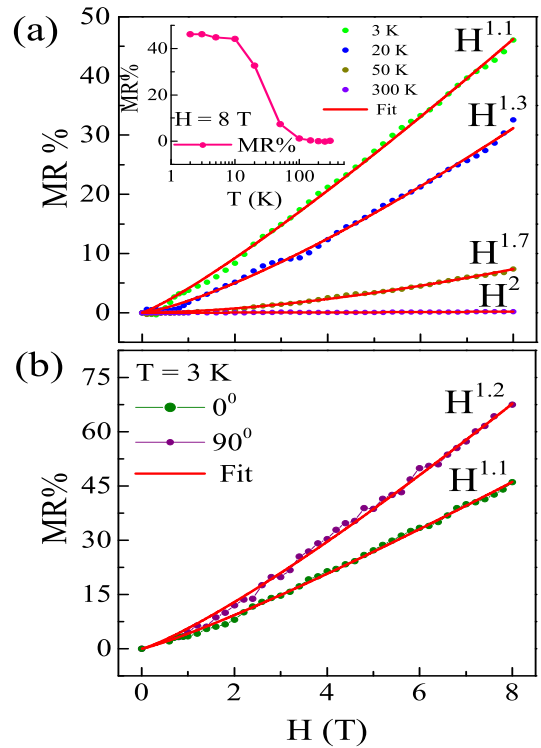}
		\caption{Magnetoresistance of Cu$_{0.05}$PdTe$_2$ (a) Longitudinal MR measured at \textit{T} = 3, 20, 50, and 300 K. The inset show temperature dependence of longitudinal MR at \textit{H} = 8 T. (b) The longitudinal and transverse MR at \textit{T} = 3 K. The inset shows the field dependence of longitudinal MR data at \textit{T} = 3 K.}
		\label{fig:Figure2}
	\end{figure}
	
	We have shown the LMR measured for temperature ranging between 3 - 300 K in Fig. \ref{fig:Figure2}(a). It is to mention that the magnetic field is applied in the direction of current ($\it{H \parallel I}$). The MR $\%$ is defined as MR = $\frac{\rho (H)-\rho (0))}{\rho (0)}$ $\times$ 100 $\%$. The LMR is positive throughout the measured temperature range and follows quadratic field dependence for $T\ge100$ K and at low temperatures, it exhibits quasi linear field dependence. This is different from parent PdTe$_2$ which shows quadratic field dependence in the whole temperature range (\textit{T} = 2 - 300 K) \cite{zheng2018detailed, meng2019planar}. The quadratic field dependence of MR in the semimetals arises due to electron-hole compensation \cite{leahy2018nonsaturating}. The intercalation of Cu in PdTe$_2$ results in dominance of p-type of carriers which is also evident from the linear field dependence of Hall resistivity in the temperature range \textit{T} = 3 - 300 K. We observed that both the longitudinal and transverse MR ($\it{H \perp I}$) exhibit unsaturated and quasi-linear field dependence at low temperatures up to 8 Tesla magnetic field (Figure \ref{fig:Figure2}(b)). This kind of unsaturated linear field dependence of MR is predicted to be due to the linear energy dispersion of Dirac fermions in quantum regime \cite{abrikosov1998quantum, abrikosov2000quantum}. However, here the critical field required for occupying the lowest Landau level in quantum regime is very high due to high carrier density. Therefore linear MR in Cu$_{0.05}$PdTe$_2$ is purely semi-classical effect and can be explained using inclusion of Cu in PdTe$_2$. Inset of Fig. \ref{fig:Figure2}(a) shows the decrease in MR values with the increase in temperature at $\it{H}$ = 8 Tesla. The LMR value of $\sim$ 46 $\%$ at \textit{H} = 8 Tesla and \textit{T} = 3 K is quite high and positive. There are various factors such as anisotropic scattering, FSA and macroscopic inhomogeneities which could lead to such positive and finite LMR \cite{hu2007nonsaturating, pippard1989magnetoresistance, jones1967magnetoresistance}. However, the high value of LMR is possibly due to FSA of Cu$_x$PdTe$_2$. It is to note that PdTe$_2$ consists multi pocket Fermi surface as seen from ARPES study \cite{yan2015identification} and intercalation of Cu would make it more complex. Therefore a finite orbital LMR would arise from those orbits of Fermi pockets which are not parallel to the direction of applied magnetic field. These pockets will have large Fermi velocity component along the direction of field and in turn give rise to finite orbital LMR.  
	
\begin{figure}[tb]
		\centering
		\includegraphics[width=\columnwidth, height=8cm]{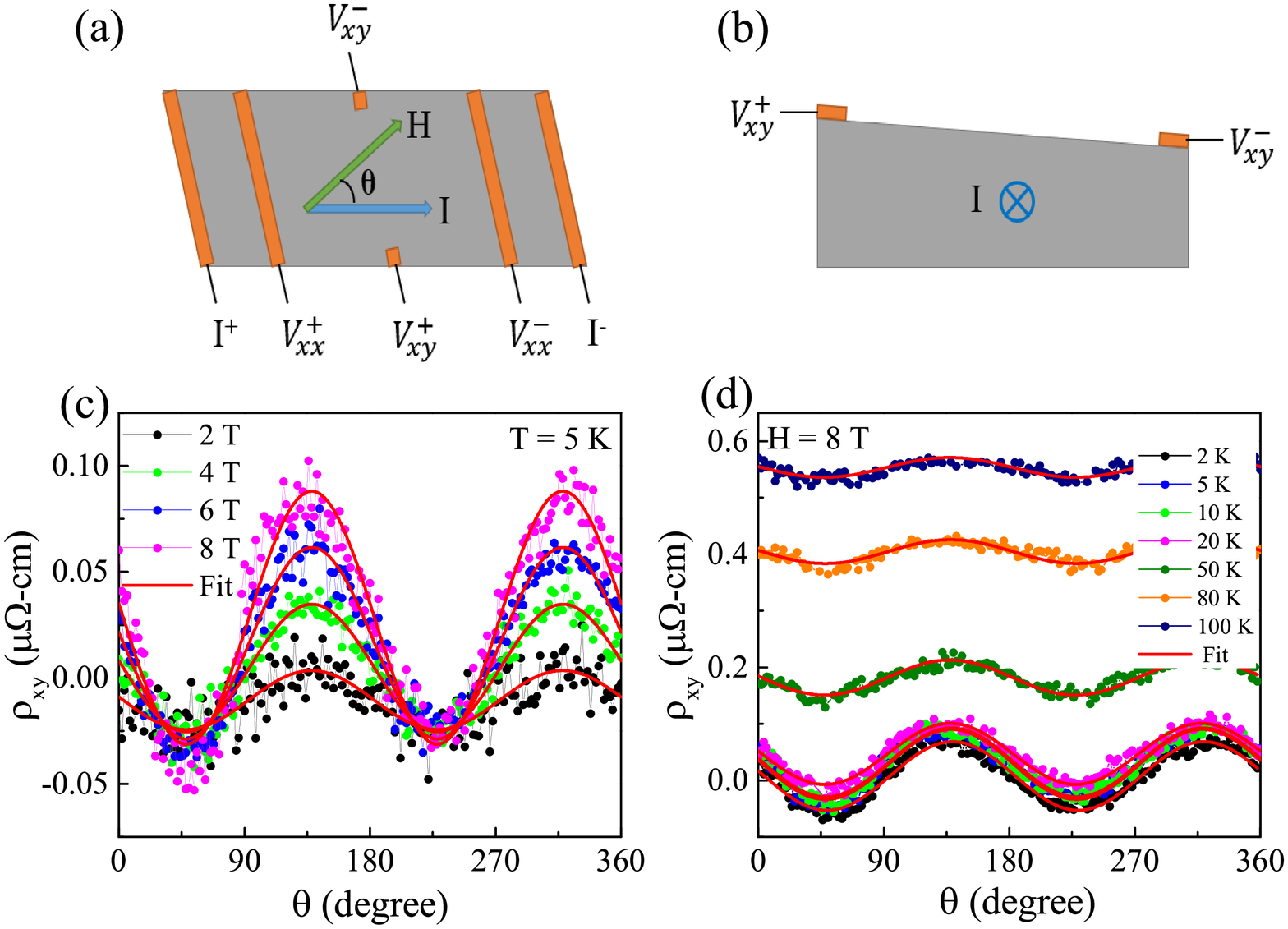}
		\caption{(a) Schematic for the PHE measurement and anisotropic longitudinal resistivity (b) Lateral view of misalignment due to non uniform thickness of sample (c) Angular dependence of the transverse component of resistivity ($\rho_{xy}$) measured in Planar Hall configuration (c) at different magnetic fields at \textit{T} = 5 K (d) at different temperatures under \textit{H} = 8 Tesla. Red curves show the fitted data.}
		\label{fig:Figure3}
	\end{figure}

	
	Figure \ref{fig:Figure3}(a) shows the schematic for  measurement of longitudinal and transverse component of resistivity in planar Hall configuration. The magnetic field is rotating in the plane of sample making an angle $\theta$ with the direction of current. The electrical current flows along the length of the sample. The voltage measured along the longitudinal direction (V$_{xx}$) gives the measure of anisotropic longitudinal resistivity whereas that along the transverse direction (V$_{xy}$) gives the planar Hall resistivity. It is to note that we have done measurement on the single crystal of dimension $\sim$ 2  $\times$ 3 $\times$ 0.3 mm$^{3}$ and the obtained value of resistance in transverse direction is as low as $\mu\Omega$, which is close to the instrumental sensitivity of the PPMS in DC constant current mode. It can be seen that although there is some noise in the data set, the angle dependence of the resistivity can be clearly discerned. We took more data points at closer angle interval in order to minimize the error in our analysis.
	
	The transverse component of resistivity ($\rho_{xy}$) can have contributions from following factors: (i) The intrinsic planar Hall component having an angular dependence of the type sin$\theta$.cos$\theta$. (ii) The misalignment of Hall contacts which results in the in-plane and out-of-plane longitudinal resistivity components with the angular dependence of the type cos$^2\theta$ and sin$^2({\theta + \delta})$. Here both cos$^2\theta$  and sin$^2({\theta+\delta}$) are symmetrical with respect to the angle $\theta$ and hence can be ignored as $\rho_{xy}$ is an odd function of $\theta$. (iii) The constant Hall resistivity component due to the non-uniform thickness of the sample (see Fig. \ref{fig:Figure3}(b)). Thus, the transverse resistivity $\rho_{xy}$ measured in the planar Hall configuration can be written as 
	\begin{equation}
		\rho_{xy} = -a. sin\theta.cos\theta + b. cos^2\theta + c
		\label{eq1}
	\end{equation}
	where \textit{a}, \textit{b}, and \textit{c} are the constants. 
	The additional factor of normal Hall resistivity arising due to any mismatch in the current and magnetic field plane can be subtracted by taking an average of the data obtained under the positive and negative magnetic field. Figure \ref{fig:Figure3}(c) and \ref{fig:Figure3}(d) represent the angular dependence of $\rho_{xy}$ under different magnetic field at $\it{T}$ = 5 K and under different temperatures at $\it{H}$ = 8 Tesla, respectively. The observed $\rho_{xy}$ exhibits a period of $\pi$ with the valley and peak at $\sim$ $\pi$/4, and $\sim$ 3$\pi$/4 respectively, which fits well (shown by red solid line) with Eq. \ref{eq1}. The parameters \textit{a}, \textit{b} and \textit{c} obtained from fitting are used to extract the planar Hall resistivity, which is defined theoretically as \cite{nandy2017chiral, burkov2017giant},
	\begin{equation}
		\rho^{PHE}_{xy} = -\Delta\rho. sin\theta. cos\theta
		\label{eq2}
	\end{equation}
	where $\Delta\rho$ = $\rho_{\perp}$ - $\rho_{\parallel}$ represents the amplitude of PHE. The $\rho_{\perp}$ and $\rho_{\parallel}$ are the resistivities for $\it{H \perp I}$ and $\it{H \parallel I}$, respectively. Generally, $\Delta\rho$ and thus PHE should vanish in this configuration (electric and magnetic field are parallel), because both $\rho_{\perp}$ and $\rho_{\parallel}$ can be considered as zero field resistivity in the absence of Lorentz force. However in topological semi-metals, $\rho_{\parallel}$ follows negative field dependence due to chiral anomaly \cite{nandy2017chiral, burkov2017giant}. This results in non-zero value of $\Delta\rho$ and hence periodic dependence in PHE and anisotropic longitudinal MR is observed.
	
	\begin{figure}[tb]
		\includegraphics[width= 8cm, height=10cm]{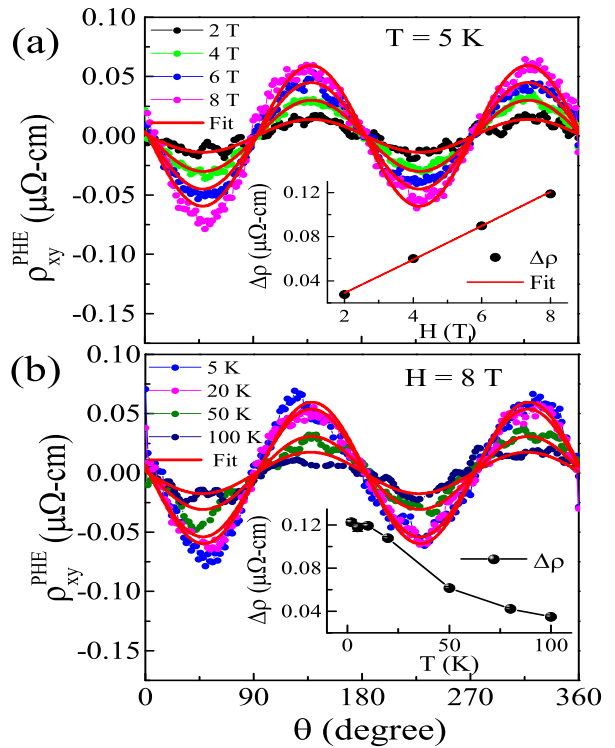}
		\caption{Planar Hall resistivity for Cu$_{0.05}$PdTe$_2$ and the amplitude of PHE ($\Delta\rho$) extracted from fitting. (a) PHE for Cu$_{0.05}$PdTe$_2$ at different magnetic fields at \textit{T} = 5K. Inset shows the field dependence of $\Delta\rho$. (b) The PHE under various temperatures at \textit{H} = 8 Tesla. Inset shows temperature dependence of $\Delta\rho$ at \textit{H} = 8 Tesla.}
		\label{fig:Figure4}
	\end{figure}
	
	The components of planar Hall resistivity at different fields for \textit{T} = 5 K and at different temperatures under $\it{H}$ = 8 Tesla field are shown in Fig. \ref{fig:Figure4}(a) and \ref{fig:Figure4}(b) respectively. The value of $\Delta\rho$ extracted by fitting the $\rho_{xy}^{PHE}$ data using Eq. \ref{eq2} are shown in the corresponding inset of these figures. For topological metals, the $\Delta\rho$ is predicted to show \textit{H}$^2$ dependence in low field region, and linear field dependence at high magnetic field \cite{burkov2017giant}. Although Chen \textit{et al.} has also discussed \textit{H}$^2$ dependence for \textit{H} $\leq$ 4 Tesla at \textit{T} = 10 K in MoTe$_{2}$, the results shown are more like linear than quadratic in the full field range \textit{H} = 0 - 8 Tesla  \cite{chen2018planar}. The observed linear field dependence of $\Delta\rho$ in our compound is similar to other semimetals like MoTe$_{2}$ and VAl$_{3}$ \cite{chen2018planar, singha2018planar}.  Here it is to mention that other effect such as FSA, carrier concentration etc. may further complicate the field dependence of $\Delta\rho$.
	
	\begin{figure}[tb]
		\includegraphics[width=\columnwidth]{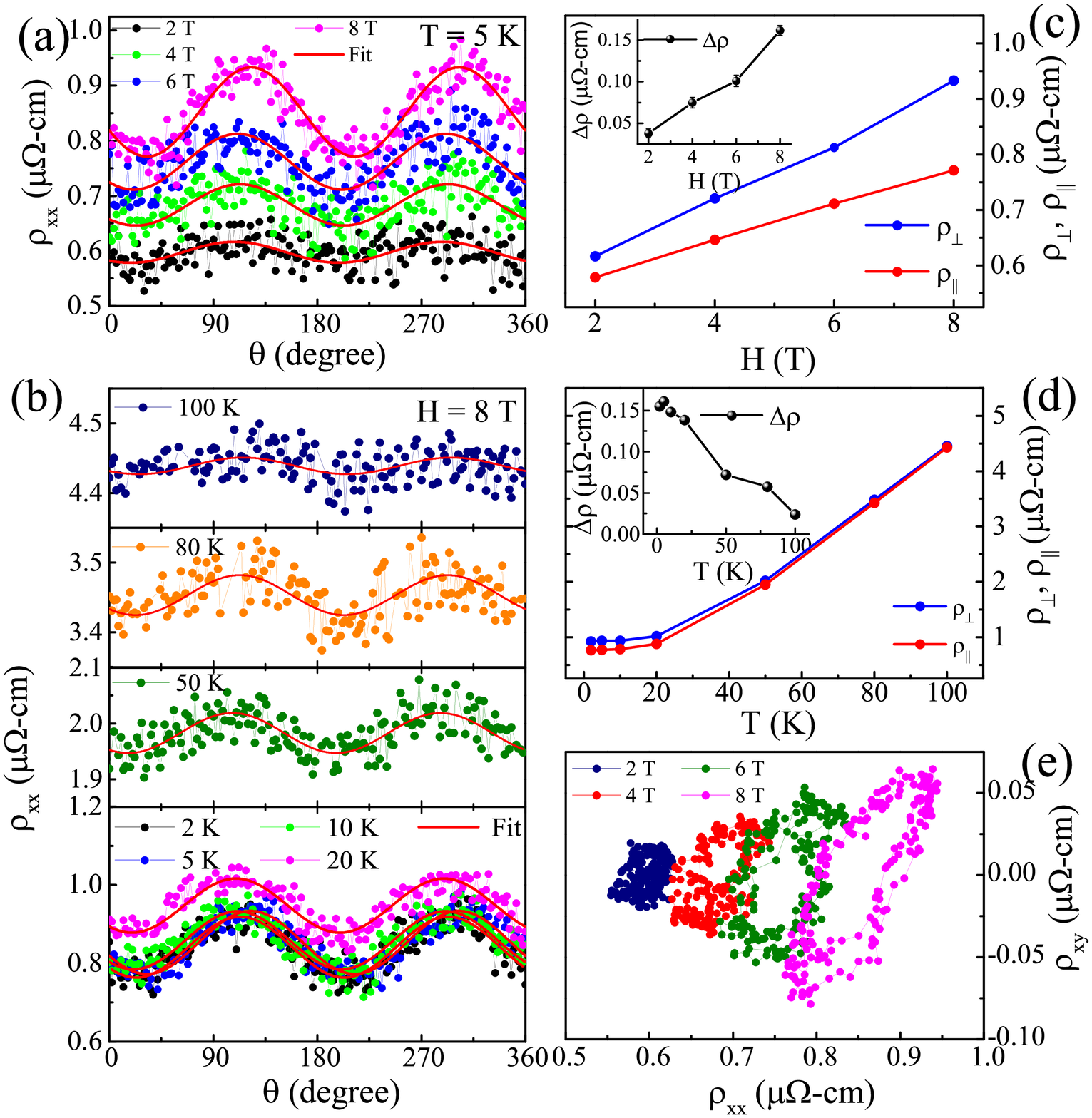}
		\caption{Anisotropic longitudinal resistivity of Cu$_{0.05}$PdTe$_2$ and the extracted resistivities $\rho_\perp$, $\rho_\parallel$ and $\Delta\rho$.  (a) The angular dependence of $\rho_{xx}$ for \textit{H} = 2, 3, 6, and 8 T at \textit{T} = 5 K. (b) Magnetic field dependence of $\rho_{\perp}$ and $\rho_{\parallel}$ at \textit{T} = 5 K. Inset shows the field dependence of $\Delta\rho$ at \textit{T} = 5 K (c) The angular dependence of $\rho_{xx}$ under various temperatures at \textit{H} = 8 T. (d) Temperature dependence of $\rho_{\perp}$ and $\rho_{\parallel}$ at \textit{H} = 8 T. Inset shows temperature dependence of $\Delta\rho$ at \textit{H} = 8 T (e) Parametric plot of the planar and AMR signals shows the orbits at \textit{T} = 5 K obtained by plotting $\rho_{xx}$ vs $\rho_{xy}$ with $\theta$ as the parameter with \textit{H} being kept fixed.}
		\label{fig:Figure5}
	\end{figure}
	
	The PHE data fits well with Eq. \ref{eq2}, where $\Delta\rho$ is assumed to be due to chiral anomaly only and hence PHE is considered an important tool to detect the Dirac (chiral) fermions \cite{nandy2017chiral}. However, it is not sufficient proof for existence of chiral anomaly \cite{kumar2017extremely}, as the materials having anisotropic magnetoresistance (AMR) are also observed to exhibit PHE \cite{nandy2017chiral}. Therefore, a robust signature of chiral anomaly would be the existence of PHE, negative LMR along with angular narrowing of planar AMR at low temperatures \cite{burkov2017giant}. Figure \ref{fig:Figure5}(a) and \ref{fig:Figure5}(b) shows AMR for different magnetic fields at $\it{T}$ = 5 K and at different temperatures in $\it{H}$ = 8 Tesla respectively. The AMR values exhibits a period of $\pi$ with maxima at $\sim$ $\pi$/2 and $\sim$ 3$\pi$/2, \textit{i.e.} when magnetic field is perpendicular to the direction of current. The AMR data is fitted using the equation\cite{nandy2017chiral, burkov2017giant}; 
	\begin{equation}
		\rho_{xx} = \rho_{\perp} - \Delta\rho. cos^2\theta.
		\label{eq3}
	\end{equation}
	The values of $\rho_{\perp}$ and $\rho_{\parallel}$ obtained from the fitting are plotted against magnetic field and temperature in Fig. \ref{fig:Figure5}(c) and \ref{fig:Figure5}(d) respectively. Figure \ref{fig:Figure5}(c) shows that both $\rho_{\perp}$ and $\rho_{\parallel}$ increase with the increase in magnetic field. The faster increase in $\rho_{\perp}$ in comparison to $\rho_{\parallel}$ results in increase of $\Delta\rho$ value with the magnetic field (inset of Fig. \ref{fig:Figure5}(c)). However, $\Delta\rho$ decreases with the increase in temperature (inset of Fig. \ref{fig:Figure5}(d)). We tried to detect the angular narrowing of AMR with decreasing temperature but we could not observe its signature quite distinctively in our data.
	
	Comparing our results for PHE and AMR values for Cu$_{0.05}$PdTe$_2$ with those for PdTe$_2$ reported in literature, we observe that the magnitudes of PHE and AMR in Cu$_{0.05}$PdTe$_2$ are comparable to that of PdTe$_2$ reported by Xu \textit{et al.} \cite{xu2018planar}. The planar AMR value is $\sim$ 10-15 $\%$ at 2 K and 8 Tesla, which is five to seven times smaller than transverse MR value of 75 $\%$ observed at same temperature and field. In such scenario, chiral anomaly contribution is obscured by orbital MR. The dominance of in-plane orbital MR over chiral anomaly for transition metal dichalcogenides has been discussed in the literature \cite{liu2019nontopological, li2020planar, meng2019planar}. However, its origin in context to PHE is not properly understood. We speculate it to be associated with FSA.
	
	We tried to fit the longitudinal magnetoconductivity data in order to see the chiral contribution to magnetoconductance. The expression for longitudinal magnetoconductivity in weak field region is given by \cite{kim2013dirac}; $\sigma_{xx}$ = $\sigma_{WAL}$.(1 + C$_W$\textit{H}$^2$) + $\sigma_n$; where $\sigma_{WAL}$ = $\sigma_0$ + p$\sqrt{H}$ ($\sigma_0$ is zero field conductivity) is the conductivity due to weak antilocalization effect and $\sigma_n^{-1}$ = $\rho_0$ + q\textit{H}$^2$ (p, q are constants and C$_W$ is the chiral coefficient, originating from the topological \textit{\textbf{E.H}} term) is the contribution from the conventional non-linear band around the Fermi level. We could fit the data quite well and the value of chiral coefficient C$_W$ lies in the range -0.011 $\pm$ 0.012 to -0.018 $\pm$ 0.015 for \textit{T} = 3 - 20 K. Although the errors in C$_W$ are comparable to the value of chiral coefficient, the total value obtained is lower than that for the systems with chiral anomaly \cite{kim2013dirac, huang2015observation}. The existence of Dirac point below Fermi level, and anisotropy in Fermi surface makes it difficult to separate the contribution of conventional fermions and Dirac fermions. In ideal case, $\rho_{\perp}$ is independent of magnetic field and $\rho_{\parallel}$ decreases with the increase in magnetic field, but experimentally obtained $\rho_{\perp}$ values are found to show some field dependence owing to the shape of the Fermi surface and limitations of single band approximation, which in turn results in finite positive orbital MR \cite{meng2019planar, liu2019nontopological}.    
	
	We further plotted the amplitude of PHE versus AMR at different magnetic field and $\theta$ as a parameter (Fig. \ref{fig:Figure5}(e)). In these parametric plots, the orbit start with small $\rho_{xx}$ value (at the left) and expands to large $\rho_{xx}$ (at the right) without showing any saturation up to \textit{H} = 8 Tesla. The shape of orbits in parametric plot is tilted prolate spheroid (ellipsoidal) which is different from the oval-shaped and shock-wave pattern of orbits reported for PtSe$_2$ and NiTe$_2$ respectively \cite{liu2019nontopological, li2020planar}. In an ideal condition if Fermi level is situated near the Dirac point, the orbit for a Dirac/Weyl point is of spherical shape. Liang \textit{et al.} has reported paraboloid and deformed oval shaped concentric orbits for Na$_3$Bi and GdPtBi respectively, which is attributed to the chiral anomaly \cite{liang2018experimental}. In case of chiral anomaly, the size of orbits enlarges more toward right on $\rho_{xx}$ axis for higher magnetic field, which has been associated with the pumping of quasi-particles from left chiral Fermi pocket to right chiral Fermi pocket. It to mention that quasi-particle scattering might balance the charge imbalance between the chiral pockets \cite{li2016chiral}. In Bi, PtSe$_2$ and NiTe$_2$, orbit starts at left and end on right exhibiting no concentric orbits, but they have different shapes owing to the different FSA in these materials \cite{liang2018experimental, liu2019nontopological, li2020planar}. The prolate spheroid shaped orbits in parametric plots for Cu$_x$PdTe$_2$ are the signature of Fermi surface and mass anisotropies, which is also evident from finite and high value of LMR. \\  
	
	\section{Conclusion}
	We have performed PHE studies on Cu$_{0.05}$PdTe$_2$. The analysis of data shows that both $\rho_{\perp}$ and $\rho_{\parallel}$ exhibit a positive field dependence and the non-zero anisotropic resistivity arises due to the fact that $\rho_{\perp}$ increases more rapidly than $\rho_{\parallel}$ with the increase of magnetic field. The Cu$_{0.05}$PdTe$_2$ shows positive LMR along with tilted prolate spheroid in parametric plot. The longitudinal magnetocoductivity data gives low value of chiral coefficient in comparison to the system with chiral anomaly and negative LMR. These observations suggests that for Type-II topological materials having positive LMR, there is need to understand the FSA to reveal the true origin of PHE. \\
	
	\textbf{Acknowledgment}: We acknowledge Advanced Material Research Center (AMRC), IIT Mandi for the experimental facilities. Sonika and Shailja acknowledge IIT Mandi for HTRA fellowship. 
	
	
	\bibliography{CuPdTe2}

\end{document}